\documentclass[mypaper,7pt,twoside]{CoAst}

\usepackage{graphicx,fancyhdr}
\renewcommand{\chaptermark}[1]{\markboth{#1}{}}

\newcommand{\kopf}{\small\itshape Comm. in Asteroseismology \\ Contribution to the Proceedings of the 38$^{th}$\,LIAC\,/\,HELAS-ESTA\,/\,BAG, 2008
}

\newcommand{\Authors}[1]{\begin{center}\normalsize\bf\sf #1 \end{center}}

\renewcommand{\author}[1]{\begin{center}\normalsize\bf\sf #1 \end{center}}
\newcommand{\Address}[1]{\begin{center}\small\sf #1 \end{center}}

\DeclareGraphicsExtensions{.eps,.jpg}

\newcommand{\Session}[1]{{\vspace{3mm}\small \noindent  \hspace*{3mm} Session: } #1 \normalsize}

	\newcommand{\one}{\small Physics and uncertainties in massive stars on the MS and close to it}

\renewenvironment{abstract}{\section*{Abstract}\normalsize\sf}{}
\newcommand{\References}[1]{\begin{flushleft}{\large References\\}\vspace*{2mm}\small #1 \end{flushleft}}

\newcommand{\chapterCoAst}[2]{\chapter[\sf\normalsize #1\\ \footnotesize \hspace*{5mm}by #2 \sf\normalsize][]{#1\\}\rhead[\fancyplain{}{\sf\footnotesize \center{#1}}]{\fancyplain{}{\sffamily\thepage}}\lhead[\fancyplain{\kopf}{\sffamily\thepage}]{\fancyplain{\kopf}{\sf\footnotesize \center{#2}}}}

\newcommand{\figureCoAst}[5]{\begin{figure}[#4]
\centering
\includegraphics*[#5]{#1}
\caption{#2}
\label{#3}
\end{figure}}

\renewcommand{\chaptername}{}

\def\rfr{\smallskip\par\noindent
        \hangindent=7truemm
        \hangafter=1}

\begin{document}
\sf

\chapterCoAst{Overshooting}{B. Dintrans}
\Authors{B. Dintrans} 
\Address{LATT, Universit\'e de Toulouse, CNRS, 14 av. Edouard Belin, 31400 Toulouse, France}

\noindent
\begin{abstract}
Overshooting occurs in stars when convective elements penetrate into
adjacent radiative zones. In the Sun, it leads to the so-called
`tachocline' at the base of the outer convection zone and this region
is becoming a key ingredient of the standard solar dynamo model as
strong toroidal magnetic fields may be generated there. However this
overshoot is not predicted by the mixing-length theory of convection
where convective elements must stop at the border of a convectively
unstable region. I will review the main properties of this convective
overshoot in stellar interiors, with in particular its subtle dependence
with the thermal diffusion, and will present the most recent results
obtained from 2-D and 3-D direct numerical simulations (DNS) of
penetrative convection.
\end{abstract}

\Session{\one} 

\section*{Introduction}

In the standard model of thermal convection based on the mixing-length
phenomenology (B\"ohm-Vitense 1958), the penetration of convective elements into surrounding stable layers is not possible. Indeed, the mixing-length velocity $v$ of a convective blob is related to its temperature contrast $\delta T$ by

\begin{equation}
v^2 \propto {\delta T\over T} g \ell \hbox{~with~} \delta T \propto \nabla - \nabla_{\hbox{\scriptsize ad}},
\end{equation}
where $g$ and $\ell$ denote the gravity and the mixing length, respectively, and $\nabla$ and $\nabla_{\hbox{\scriptsize ad}}$ the usual temperature `nabla'. Following Schwarzschild's criterion, the interface that separates a convective zone from a radiative one corresponds to $\nabla = \nabla_{ad}$ therefore the blob must stop here. However, the mixing-length formalism is based on an acceleration budget and setting the blob acceleration to zero does not mean that its velocity is itself zero, i.e., the mixing-length theory neglects {\it inertia}. In fact, the blob penetrates into the stable layer over a small distance, then it is slowed down by the buoyancy braking and finally stops. This small penetration is not  anecdotal because it has important consequences in stellar physics:

\noindent - it leads to different evolutionary tracks in the HR diagram as massive stars live longer on the main sequence due to the additional mixing induced by the overshooting above the convective core (Rosvick \& Vandenberg 1998, Perryman et al. 1998). 

\noindent - in the standard $\alpha-\Omega$ model of the solar dynamo, the regeneration of poloidal magnetic fields into toroidal ones takes place in the tachocline where vertical and latitudinal differential rotation is expected to be most efficient (Parker 1955, Brandenburg \& Subramanian 2005).

\noindent - heliosismology showed that the solar core rotation is almost rigid (Brown et al. 1989). Internal gravity waves propagating in stellar radiative zones are suspected to play the main role in this angular momentum redistribution (Zahn et al. 1997, Talon \& Charbonnel 2003). The most wide-spread excitation model involves penetrative convection from neighboring convection zones as strong downdrafts extend a substantial distance into the adjacent stable zones so that internal gravity waves can be randomly generated (Dintrans et al. 2005).

\section*{Modelling of overshooting}

Assuming that the temperature gradient in the overshoot layer is close to the adiabatic one of the next convection zone, Roxburgh (1978) investigated the overshoot of convective cores and derived what it is now called the `Roxburgh's integral constraint': 

\begin{equation}
\int^{r_c}_0 (L-L_{\hbox{\scriptsize rad}}) d \left( {1\over T}\right) > 0,
\end{equation}
where the radius $r_c$ corresponds to the edge of the overshoot region, while $L$ and $L_{\hbox{\scriptsize rad}}$ denote the total and radiative luminosity, respectively. The overshoot layer corresponds in this relation to the region where $L<L_{\hbox{\scriptsize rad}}$ (i.e., negative contribution) while the convection zone means $L>L_{\hbox{\scriptsize rad}}$ (i.e., positive contribution). However, this relation gives an upper limit of the overshoot extent because it does not take into account the dissipation acting on the penetrating elements. Indeed, the viscous dissipation ${\cal D}_{\nu}$ enters in the heat equation and affects the blob dynamics such that the correct relation reads now (Roxburgh 1989):

\begin{equation}
\int^{r_c}_0 (L-L_{\hbox{\scriptsize rad}}) d \left( {1\over T}\right) = \int^{r_c}_0 4\pi r^2 {{\cal D}_{\nu}\over T} dr\hbox{~with~} {\cal D}_{\nu} \sim \nu \left( {\partial u_i \over \partial x_j}\right)^2,
\label{rox89}
\end{equation}
where $\nu$ denotes the kinematic viscosity. As the viscous dissipation is always positive, this term acts to decrease the negative part of the integrand in the left-hand side and thus the radius $r_c$.

Following Roxburgh's investigation, Zahn (1991) showed that the penetration extent also depends on the value of the Peclet number $Pe$ associated with the convective elements, that is, 

\begin{equation}
Pe  = {v \ell \over \chi},
\end{equation}
where $\chi$ denotes the thermal diffusivity. Small Peclet's numbers mean that the radiative diffusion is stronger than the advection and the convective blob rapidly loses its identity compared to the surrounding medium. As a consequence, it does not feel any more the buoyancy braking and travels over long distances in the radiative zone (Zahn called this regime `overshoot'). On the other hand, a blob with a large Peclet number does not thermalize rapidly with its surrounding (i.e., no radiative losses) and keep its identity: it is then strongly slowed down by the buoyant force and its penetration is weaker (the so-called `penetration' regime). Penetration leads to an adiabatic stratification below the convection zone because of the induced entropy mixing, whereas overshoot has no significant influence on the local stratification.

Assuming that the velocity $W$ of the penetrating motions obeys the usual mixing-length scaling of thermal convection, Zahn derived the following relations for both the subadiabatic extent of a convective region and the width of the thermal boundary layer:

\begin{equation}
L_p \propto f^{1/2} W^{3/2} \hbox{~and~} L_{\hbox{\scriptsize bound}}\propto \left( \chi H_p\over g\right)^{1/2},
\label{JPZ91}
\end{equation}
where $f$ denotes a filling factor of convective plumes (i.e., fraction of the area occupied by the plumes) and $H_p$ the pressure scale-height at the bottom of the convection zone. Applied to the Sun, these two scalings lead to a penetration extent of about $0.2-0.3H_p$ ($\sim 15\ 000$ km) and an overshoot layer $\simeq 1$ km. These scalings seem however to overestimate the amount of convective penetration inferred by helioseismology as the study of the oscillations in the asymptotic laws of solar acoustic modes rather predicts an upper limit of $\sim 0.1 H_p$ for the extent of the whole convective overshooting (Roxburgh \& Vorontsov 1994; Christensen-Dalsgaard et al. 1995; see also Aerts et al. 2003 for the case of a massive star).

\section*{Numerical simulations of overshooting}

\subsection*{The bad news}

The rapid development of supercomputers has lead several groups to undertake direct numerical simulations of convective overshooting. However, many numerical difficulties appear when one tries to simulate stellar convection:

\noindent - the molecular viscosity in stellar interiors is small, typically $\nu \sim 1$ cm$^2$/s (Parker 1979). Granulation cells in the upper convection zone of the Sun correspond to a typical size and velocity of about $L=1000$ km and $U=1$ km/s, respectively. It leads to Reynolds' number of about $Re = U L / \nu \sim 10^{11}$, meaning that solar convection is a highly turbulent and nonlinear phenomenon.

\noindent - we know from Kolmogorov's theory of turbulence (Kolmogorov 1941) that
the ratio between the largest scale (i.e., the injection scale $L$) and the smallest one (i.e., the dissipation scale $\ell_d$) behaves as $L/\ell_d \sim Re^{3/4}$ such that the solar dissipation scale is about one centimeter. In other words, one needs (at least) $10^8$ gridpoints in each spatial direction to well reproduce both the injection of energy in the largest structures and its viscous dissipation in the smallest ones (the turbulent cascade scenario). As this number of gridpoints is clearly beyond the scope of today supercomputers, several numerical techniques have been developed to overcome this restriction: (i) instead of doing Direct Numerical Simulations (DNS), one can perform Large-Eddy Simulations (LES) where the small-scale dynamics is modelled using a sub-grid scale approach (Lesieur 2008); (ii) one can also artificially increase the dissipation scale by adding a large-scale viscosity in the whole computational domain, the price to pay being that motions are essentially laminar ($Re\sim 10^2\cdots 10^4$).

\noindent - finally, another problem concerns the different time scales that coexist in a convection zone. Indeed, the local turnover time scale of convection is roughly given by:

\begin{equation}
\tau \sim H_p / U_{RMS},
\end{equation}
where $U_{RMS}$ is the typical (turbulent) RMS velocity of convective eddies. Applied to the solar convection zone, it leads to turnover time scales of about the day at the surface to several months at the bottom (Spruit 1974). As a consequence, simulations should reproduce motions whose dynamical time scales can vary by two orders of magnitude in the computational domain. Moreover, the thermal time scale is also much more longer than these turnover time scales by orders of magnitude as, e.g., the solar thermal flux expressed in dynamical units is very small at the bottom of the convection zone ($F_\odot / (\rho c^3_s) \sim 10^{-11}$). This ratio is in essence equal to the ratio of the dynamical to thermal time scales, meaning that realistic simulations should be integrated on a very long time to make sure that a thermal-relaxed state is reached. It is not possible in practice to do that with a decent spatial resolution and the thermal fluxes are commonly overestimated by orders of magnitude in numerical simulations.

\begin{figure}\centerline{
\includegraphics[width=5cm]{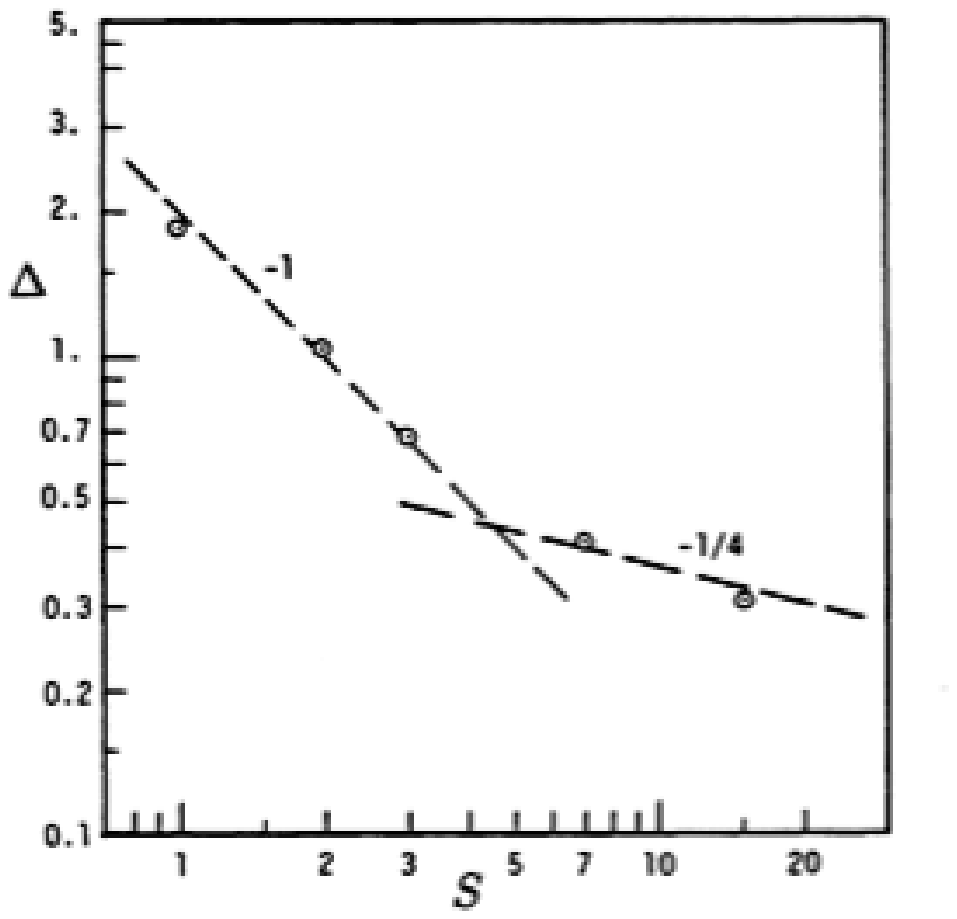}
\includegraphics[width=6cm]{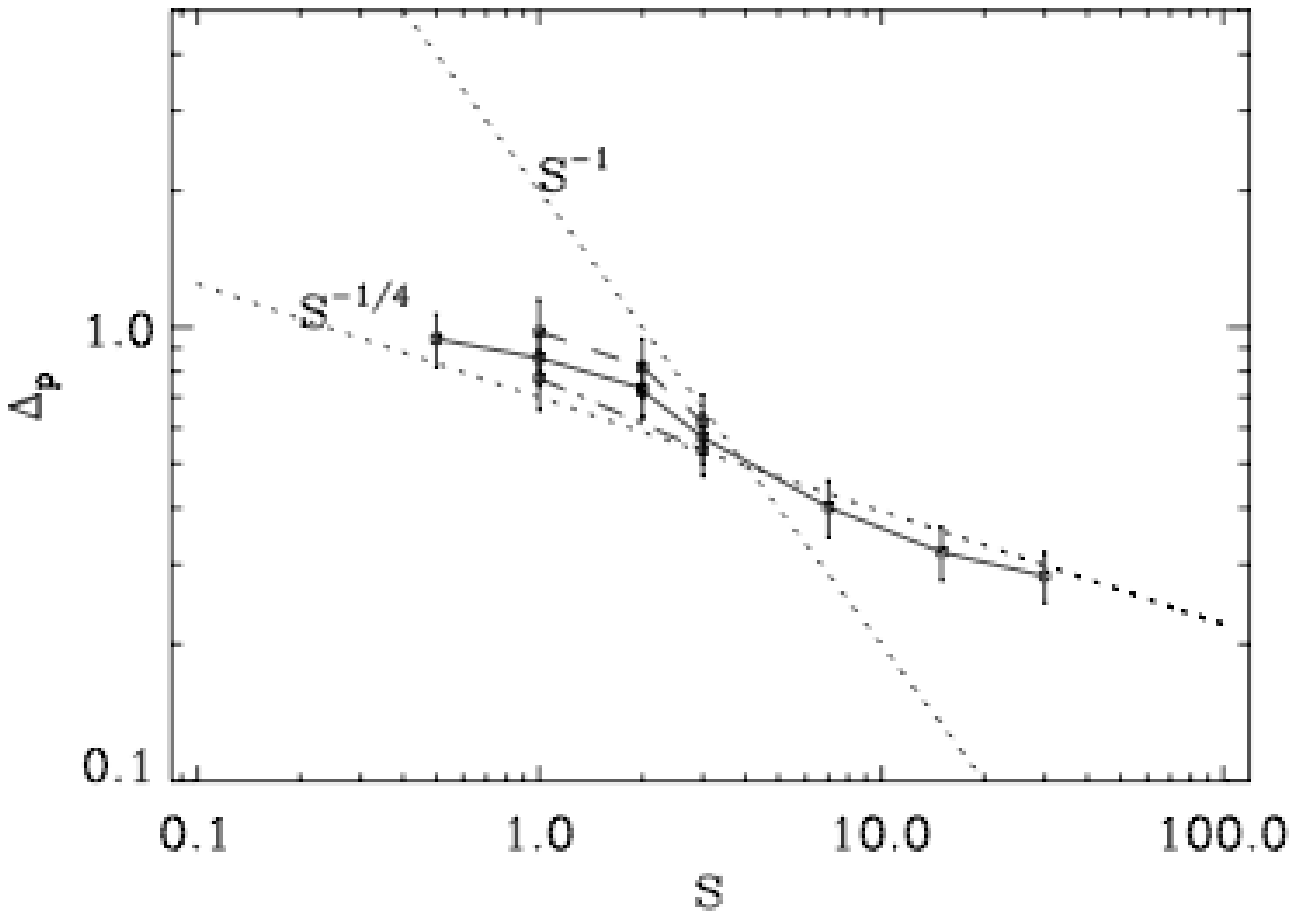}}
\caption{Penetration extent $\Delta$ v.s. stability parameter $S$. \textit{Left}: in the 2-D case where both the penetration and overshoot regimes exist (from Hurlburt et al. 1994). \textit{Right}: in the 3-D case where only the overshoot regime remains (from Brummell et al. 2002).}
\label{stab}
\end{figure}

\subsection*{DNS of convective overshooting in local cartesian boxes}

Following Hurlburt et al. (1984), the majority of DNS done so far use polytropic solutions for the internal structure of the star. Indeed, polytropes with $\rho \propto T^m$ ($m$ being the polytropic index) allow to easily specify the strength of the convective instability as the hydrostatic equilibrium implies in that case

\begin{equation}
{dT\over dz} = - {g\over (m+1) R_*},
\end{equation}
where $g$ is the downward-directed gravity, $z$ the altitude and $R_*$ the perfect gas constant. For a monatomic gas with an adiabatic index $\gamma = 5/3$, this relation shows that the temperature gradient is equal to the adiabatic one $(=-g/c_p)$ when the polytropic index  is $m_{ad} = 1/(\gamma - 1) = 3/2$. As a consequence, a stable layer corresponds to $m>m_{ad}$ and a convectively-unstable one to $m<m_{ad}$.

Hurlburt et al. (1986, 1994) used this model to study the overshooting in 2-D simulations of compressible convection for a solar case, that is, a radiative zone with polytropic index $m_3 > 3/2$ located above a convective zone with polytropic index $m_2 < 3/2$. In particular, they studied the influence of the relative stability parameter $S$ on the penetration extent $\Delta$, where $S = - (m_{\hbox{\scriptsize ad}}-m_3)/(m_{\hbox{\scriptsize ad}}-m_2)$, and found the following scalings (Fig.~\ref{stab}, left):

\begin{equation}\left\{ \begin{array}{l}
\Delta \propto S^{-1} \hbox{~for small S-values: penetration regime},\\ \\
\Delta \propto S^{-1/4} \hbox{~for large S-values: overshoot regime.}
\end{array}\right.
\end{equation}
However, these scaling laws have not been fully confirmed in the 3-D case by Brummell et al. (2002) as only the overshoot regime $\Delta \propto S^{-1/4}$ is recovered (Fig.\ref{stab}, right). The reason simply lies in the smaller filling factor $f$ for the convective plumes compared to the 2-D case (Eq.~\ref{JPZ91}): the spaced downdrafts that penetrate into the stably stratified layer are not enough to maintain an adiabatic stratification below the convection zone and only the thermal boundary layer exists. 

\figureCoAst{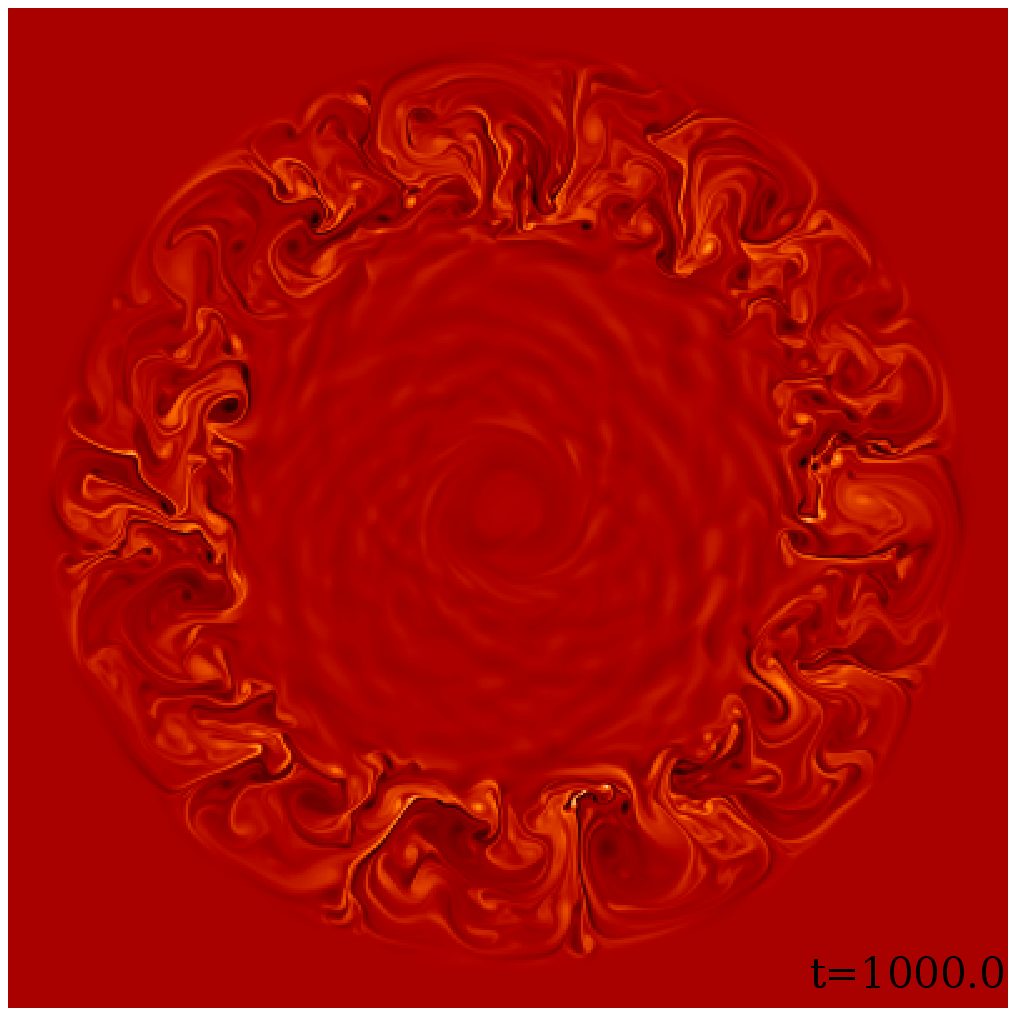}{Vorticity isocontours in a 2-D star-in-a-box simulation of penetrative convection for a solar-like star (Dintrans \& Brandenburg 2009).}{starinbox}{!h}{clip,angle=0,width=6cm}

Another interesting 2-D and 3-D simulations of penetrative convection are the ones done by Roxburgh \& Simmons (1993) and Singh et al. (1998). Indeed, a more realistic model of the star interior has been considered in the 2-D case through a temperature-dependent radiative conductivity. The main result is that the viscous dissipation contribution in the RHS of Roxburgh's integral constraint (\ref{rox89}) decreases with the Prandtl number. It suggests that in the astrophysical limit of very small Prandtl numbers ($Pr\sim 10^{-6}$), the stable and unstable contributions in the LHS of Roxburgh's integral strictly balance each other and are sufficient to obtain a good estimate for the overshoot extent (Roxburgh \& Simmons 1993). In the 3-D case (Singh et al. 1998), a sub-grid scale modelling of the small scales led to the confirmation of the scaling relationships between this extent and both the imposed bottom flux and the vertical velocity of convective downdrafts (see Eq.~\ref{JPZ91}).

\subsection*{Global DNS}

In the last decade, global direct simulations of overshooting have greatly improved following two numerical approaches: 

\noindent - the spherical shape of the star is reproduced using a pseudo-spectral code where physical fields are projected onto spherical harmonics. Browning et al. (2004) used this method with the 3-D anelastic ASH code to study the core overshooting in a massive A-type star of $2M_\odot$, where only the inner 30\% by radius of the star has been considered (convective core and some of its surrounding radiative envelope). Although the achievement of the thermal relaxation in these simulations remains questioning, they found that the convective core adopts a prolate shape while overshooting is larger at the equator than at the poles, yielding an overall spherical shape to the whole central region.

\noindent - the star is embedded in a topologically Cartesian domain and the sphericity is reproduced through radial-dependent forcing or damping profiles. This novel approach has been developed by Dintrans \& Brandenburg (2009) and first results in 2-D are promising as both the penetration and overshoot regimes are observed, whereas internal gravity waves are efficiently excited in the central radiative zone (Fig.~\ref{starinbox}).

\subsection*{Conclusion}

Convective overshooting is today included in stellar evolution codes and evolutionary tracks of massive stars reproduce satisfactorily the observations of clusters when an overshooting above convective cores is assumed. Theoretical studies showed that the overshoot region is composed of two nested layers whose the size depends on the value of the Peclet number of penetrating motions. Numerical studies in 2-D and 3-D of penetrative convection confirmed this picture, except that only the overshoot layer (i.e., the thermal boundary one) exists in 3-D due to the weak filling factor of convective plumes.
These local and global direct numerical simulations are however still quite far from realistic values of stellar interiors. Flows are less turbulent than in reality and the overshoot regime dominates. Further progress are clearly needed in the numerical side of this problem (e.g., development of large-eddy simulations), in its theoretical modelling (e.g., turbulent closure models; Kupka \& Montgomery 2002) and in the observational signatures of overshooting from asteroseismology.

\References{
\rfr Aerts, C., Thoul, A., Daszy\'nska, J., et al. 2003, Sci, 300, 1926
\rfr B\"ohm-Vitense, E. 1958, ZA, 46, 108
\rfr Brandenburg, A., \& Subramanian, K. 2005, PhR, 417, 1
\rfr Brown, T.~M., Christensen-Dalsgaard, J., Dziembowski, W.A., et al. 1989, SoPh, 343, 526
\rfr Browning, M.~K., Brun, A.~S., \& Toomre, J. 2004, ApJ, 601, 512
\rfr Brummell, N.~H., Clune, T.~L., \& Toomre, J. 2002, ApJ, 570, 825
\rfr Christensen-Dalsgaard, J., Monteiro, M.~J.~P.~F.~G., \& Thompson, M.~J. 1995, MNRAS, 276, 283
\rfr Dintrans, B., Brandenburg, A., Nordlund, \AA., et al. 2005, A\&A, 438, 365
\rfr Dintrans, B., \& Brandenburg, A. 2009, \AA., in preparation
\rfr Hurlburt, N.~E., Toomre, J., \& Massaguer, J.~M. 1984, ApJ, 282, 557
\rfr Hurlburt, N.~E., Toomre, J., \& Massaguer, J.~M. 1986, ApJ, 311, 563
\rfr Hurlburt, N.~E., Toomre, J., Massaguer, J.~M., \& Zahn, J.-P. 1994, ApJ, 421, 245
\rfr Kolmogorov, A.~N. 1941, DoSSR, 30, 301
\rfr Kupka, F., \& Montgomery, M.~H. 2002, MNRAS, 330, L6
\rfr Lesieur, M. 2008, Turbulence in Fluids (Springer-Verlag)
\rfr Parker, E.~N. 1955, ApJ, 122, 293
\rfr Parker, E.~N., 1979, Cosmical Magnetic Fields, (Oxford University Press)
\rfr Perryman, M.~A.~C., Brown, A.~G.~A., Lebreton, Y., et al. 1998, A\&A, 331, 81
\rfr Rosvick, J.~M., \& Vandenberg, D.~A. 1998, AJ, 115, 1516
\rfr Roxburgh, I.~W. 1978, A\&A, 65, 281
\rfr Roxburgh, I.~W. 1989, A\&A, 211, 361
\rfr Roxburgh, I.~W. \& Simmons, J. 1993, A\&A, 277, 93
\rfr Roxburgh, I.~W., \& Vorontsov, S.~V. 1994, MNRAS, 268, 880
\rfr Singh, H.~P., Roxburgh, I.~W., \& Chan, K.~L. 1998, A\&A, 340, 178
\rfr Spruit, H.~C. 1974, SoPh, 34, 277
\rfr Talon, S., \& Charbonnel, C. 2003, A\&A, 405, 1025
\rfr Zahn, J.-P. 1991, A\&A, 252, 179
\rfr Zahn, J.-P., Talon, S., \& Matias, J. 1997, A\&A, 322, 320
}

\end{document}